\documentclass[11pt]{article}
\usepackage[pagebackref,raiselinks,hyperindex]{hyperref}

\usepackage{enumerate}
\usepackage{amssymb}
\usepackage{theorem}
\usepackage{exscale}
\usepackage{amsmath}
\newcommand{\be}{\begin{equation}}
\newcommand{\ee}{\end{equation}}
\newcommand{\bea}{\begin{eqnarray*}}
\newcommand{\eea}{\end{eqnarray*}}
\newcommand{\beq}{\begin{eqnarray}}
\newcommand{\eeq}{\end{eqnarray}}
\newcommand{\RR}{\mathbb{R}}
\newcommand{\NN}{\mathbb{N}}
\newcommand{\ZZ}{\mathbb{Z}}
\newcommand{\EE}{\mathbb{E}}
\newcommand{\PP}{\mathbb{P}}
\renewcommand{\r}{\right}
\renewcommand{\l}{\left}
\newcommand{\iznad}[2]{\genfrac{}{}{0pt}{}{#1}{#2}}
\newcommand{\diam}{\mathop{\mathrm{diam}}}
\newcommand{\supp}{\mathop{\mathrm{supp}}}
\newcommand{\Id}{{\mathop{\mathrm{Id}}}}
\newcommand{\Tr}{\mathop{\mathrm{Tr}}}
\newtheorem{thm}{Theorem}

\newtheorem{prp}[thm]{Proposition}

\theorembodyfont{\normalfont}
\newtheorem{bem}[thm]{Remark}
\newtheorem{nta}[thm]{Notation}
\newtheorem{bsp}[thm]{Example}

\begin{document}

\title{Wegner estimate and the density of states\\  of some indefinite alloy type\\
  Schr\"odinger Operators}

\author{{Ivan Veseli\'{c}}\\
{\small Fakult\"at f\"ur Mathematik,}
{\small Ruhr-Universit\"at Bochum, Germany}\\
{\small\ttfamily\href{http://www.ruhr-uni-bochum.de/mathphys/ivan/}{http://www.ruhr-uni-bochum.de/mathphys/ivan/}}
}

\date{}
\maketitle

\begin{abstract}
We study Schr\"odinger operators with a random potential of
alloy type. The single site potentials are allowed to change sign.
For a certain class of them we prove a Wegner estimate. This is a key ingredient in an existence proof of
pure point spectrum of the considered random Schr\"odinger operators.
Our estimate is valid for all bounded energy intervals and all space
dimensions and implies the existence of the density of states.
\end{abstract}

{{\bfseries Keywords:} density of states, random
  Schr\"{o}dinger operators, Wegner estimate, multi scale analysis,
  localization, indefinite single site potential}

{For somewhat different versions see: \\
\href{http://www.ma.utexas.edu/mp_arc-bin/mpa?yn=00-373}{http://www.ma.utexas.edu/mp\_arc-bin/mpa?yn=00-373}
\\ and \href{http://dx.doi.org/10.1023/A:1015580402816}{Letters in Mathematical Physics 59 (3): 199-214, 2002}.}

\section{Alloy type models and Wegner's estimate}

The subject matter of this work are families of random Schr\"{o}dinger
operators $\{ H_\omega \}_{ \omega \in \Omega} $ acting on $L^2(\RR^d)$. They have been introduced as
quantum  mechanical models  for disordered media in solid state
physics. The random Schr\"odinger operator we consider is of {\em Anderson}
or {\em alloy} type
\begin{equation}
\label{model}
H_\omega = - \Delta + V_0 +  V_\omega ,
\end{equation}
where  the negative Laplace operator $-\Delta $ corresponds to the kinetic energy, $V_0$ is a bounded
$\ZZ^d$-periodic potential and $V_\omega$ is the
random potential given by the stochastic process
\begin{equation}
\label{potential}
V_\omega (x) = \sum_{k \in \ZZ^d} \omega_k \,  u(x - k)\
 .
\end{equation}
The function $ u \colon \RR^d \to \RR$ is called 
{\em single site potential} and represents the contribution to $V_\omega$
due to a single nucleus or ion situated at a lattice point
$k \in \ZZ^d$. We assume that $u \in L^p(\RR^d)$ with $p=2$ for $d\le 3$ and $p >d/2$ for $d\ge 4$ is  compactly supported.  
The $\omega_k$  are  real-valued, random {\em coupling constants}. 
I.e.~while we fix the shape of the single site potential at each 
$k \in \ZZ^d$, its strength is allowed to vary randomly.
The random variables $\omega_k, k \in \ZZ^d$ are
independent and identically distributed  and the distribution 
measure $\mu$ of $\omega_0$ has a density $f $.
We consider the coupling constants as components of a random vector
$ \omega := \{ \omega_k \}_{ k \in \ZZ^d} \in \Omega := \times_{ k \in
  \ZZ^d} \RR$. The  probability space $\Omega$ is equipped with
the product measure $\PP := \bigotimes_{ k \in  \ZZ^d} \mu $. The corresponding expectation is denoted by $\EE$.
\medskip

To state our main technical result we introduce auxiliary objects associated to finite cubes.
We denote by $\Lambda_l$ the cube of side length $l$ and centre at
$0$ and with $H_\omega^l$ the restriction of $H_\omega$ to $\Lambda_l$ with
periodic boundary conditions (b.c.). Note that $H_\omega^l$ has purely discrete 
spectrum so we can enumerate its eigenvalues $\lambda_i (H_\omega^l)$ 
in non-decreasing order and counting multiplicities. For a bounded interval $I \subset \RR$ the spectral projection $P_\omega^l(I)$ of $H_\omega^l$ has a finite trace.

\begin{thm}[Wegner estimate]
\label{theorem1}
Let the density $f$ have compact support and  be in the Sobolev space  $ W_1^1(\RR)$ and the single site potential be of {\em generalized step function} form:
\begin{equation}
\label{u}
u(x) = \sum_{k \in \Gamma} \alpha_k \ w(x -k), \  \Gamma \subset \ZZ^d,
\end{equation}
where $ w \ge \kappa \chi_{[0,1]^d}$ with some positive $\kappa$ and $w \in L^p(\RR^d)$ with $p=2$ for $d\le 3$ and $p >d/2$ for $d\ge 4$. We assume that $ \Gamma $ is a finite set and 
\be
\label{alphastern}
\alpha^* = \sum_{k \neq 0} | \alpha_k | < |\alpha_0|.
\ee
Then we have for all $E \in \RR$
\begin{equation}
\label{resultat1}
\EE \l [ \Tr P_\omega^l( \l [E -\epsilon,E \r ]) \r ] 
\le \mbox{const } \epsilon \ l^d , \quad \forall \, \epsilon \ge 0.
\end{equation}
The constant depends on $E$ but not on $\epsilon$.
\end{thm}

The theorem remains true if we replace the periodic b.c.~by Dirichlet or Neumann ones. We call $\alpha$ the {\em convolution vector}. 

\begin{bem}
\label{Serra}
For our proof it is essential that the single site potential $u$ is of generalized step function form, since this enables us to work simultaneously with two different representations of the random potential, cf.~Section \ref{trans}. Condition \eqref{alphastern} ensures the invertibility of the block-Toeplitz operator generated by the convolution vector $\alpha$ and moreover a uniform bound on the norms of the inverses of finite truncations of this Toeplitz operator. This uniform invertibility could be alternatively ensured by an appropriate condition on the symbol of the Toeplitz operator, cf.~e.g.~Chapter 7 of \cite{BoettcherS-1990} or  Chapters 2 and 6 of \cite{BoettcherS-1999}. This will be discussed elsewhere, as announced in \cite{KostrykinV-2001}. Recently there has been increased interest in conditions on the symbol of the Toeplitz operator, which ensure merely that the norms of the inverses of finite size truncations grow at most polynomially in the size of the truncation, cf.~\cite{Serra-1998a,Serra-1996,Serra-1994,Serra-1998b,BoettcherG-1998}. 
Such conditions could be useful for the derivation of a Wegner estimate which, in turn, can be used as an ingredient od a proof of localization, although it is not sufficient to ensure the existence of the density of states, cf.~Section \ref{loc}. 
\end{bem}

In the next section we deduce the existence of the density of states from the Wegner estimate in Theorem \ref{theorem1} and discuss its role for the proof of localization. Furthermore we review earlier results for indefinite alloy type models.
Sections 3 to 5
contain the proof of the main  technical Theorem \ref{theorem1}  and the last two sections are
devoted to the discussion of generalizations of the results and the application to localization for indefinite models.
\vspace{1em}

{\bfseries Acknowledgements:}

The author is grateful for stimulating discussions with 
W.~Kirsch, K.~Veseli\'c, S.~B\"ocker, E.~Giere, N.~Minami and C.~Riebling as well as for suggestions for the improvement of an earlier version \cite{Veselic-2000b} of this work by T.~Hupfer and S.~Warzel. 
He would like to thank the  SFB 237:~``Unordnung und gro\ss e Fluktuationen'', the Ruth-und-Gert-Massenberg-Stiftung, both Germany, the MaPhySto Centre, Denmark, and the Japanese Society for the Promotion of Science  for financial support.

\section{Density of states and  localization}
\label{loc}
Under our assumptions the family $H_\omega, \omega \in \Omega$ fits into the
general theory of ergodic random Schr\"odinger operators
\cite{Kirsch-89a,CarmonaL-90,PasturF-92}. We infer two central results
from this theory.
\begin{enumerate}[(A)]
\item
\label{genthm1}
The spectrum of the family $H_\omega, \omega \in \Omega$ is non-random in the
following sense. There exists a subset $\Sigma$ of the real line and an
$\Omega' \subset \Omega$, $ \PP (\Omega') =1 $ such that for all 
$ \omega \in \Omega' $ one has
$
\sigma (H_\omega) = \Sigma
$.
The analogous statement holds true for the essential, discrete,
continuous, absolutely continuous, singular continuous, and pure 
point part of the spectrum. Note that the pure point spectrum
$\sigma_{pp}$ is the closure of the set of eigenvalues of $H_\omega$.
\item
\label{genthm2}
There exists a {\em self averaging} integrated density of states 
 associated with the family $H_\omega, \omega \in \Omega$. 
This means that the normalized eigenvalue counting functions
\begin{equation}
N_\omega^l (E) = l^{-d} \# \{ i | \ \lambda_i (H_\omega^l) < E \}
= l^{-d} \Tr P_\omega^l(]-\infty,E[) 
\end{equation}
of $H_\omega^l$ converge for almost all $\omega$ to a limit
$N := \lim_{ l \to \infty} N_\omega^l $ which is $\omega$-independent.
\end{enumerate}

We call $N$ the {\em integrated density of states} (IDS) of $H_\omega$ and
$N_\omega^l$ the {\em finite volume} IDS of $H_\omega^l$.

\begin{bem}
\label{questions}
While the two above facts (\ref{genthm1}) and (\ref{genthm2}) follow from the
general theory, one is interested in more detailed spectral properties of
specific models $H_\omega, \omega \in \Omega$, e.g.:
\begin{itemize}
\item
Which spectral types can occur in $ \sigma ( H_\omega) $? 
\item
Can something be said about the regularity of the IDS $N$ as a
function of the energy $E$? Is it H\"older continuous or does even its derivative, the {\em density of states} exist.
\end{itemize}
\end{bem}

Our result on the regularity of the IDS is strong enough to imply the existence of the density of states:

\begin{thm}[Density of states]
\label{theorem2}
Under the assumptions of Theorem \ref{theorem1} the IDS of the alloy type model
 $\{ H_\omega \}_{ \omega \in \Omega} $ is Lipschitz continuous: for all $E \in \RR$ there exists a constant $C$ such that 
 \begin{equation}
N(E) - N(E-\epsilon) \le C \, \epsilon, \quad \forall \ \epsilon >0
\ .
\end{equation} 
By Rademacher's theorem it follows that the derivative $\frac{dN}{dE}$ exists
for almost all $E$. 
\end{thm}
The analog result for $u\ge \kappa \chi_{[0,1]^d}, \kappa>0$ is proved in \cite{CombesH-94b}, cf.~also Section \ref{Botd}.
\begin{bem} 
The theorem  follows directly from (\ref{resultat1}) and the self averaging property $N (\cdot) =\EE N(\cdot)$. There is a explicit upper bound for the density of states, see (\ref{fvdos}). 
\end{bem}

The second question of Remark \ref{questions} is related to the  transport properties of
the medium modelled by $H_\omega$. A perfect crystal is described by a
Schr\"odinger operator with periodic potential. It has purely absolutely
continuous spectrum, which reflects its good electric transport
properties. In contrast to this, it has been proven that random
perturbations of this regular structure give rise to energy intervals
with pure point spectrum. This corresponds to the less effective
transport properties of random media. The existence of pure point
spectrum in this context is called {\em (Anderson) localization}.

Now we  indicate the general scheme of the proof of localization and where the Wegner estimate enters. In section \ref{sec-localization} we show  that Theorem \ref{theorem1} implies localization  for  some alloy type models with single site potentials that change sign.

A powerful tool for proving localization is the so-called
{\em multi scale analysis} (MSA), an induction argument over
increasing length scales $l_k, k \in \NN $. This technique was first applied by
Fr\"ohlich and Spencer \cite{FroehlichS-83} to the discretization of
the Schr\"odinger operator (\ref{model}) and underwent since then a
number of strengthenings \cite{MartinelliS-1985b,FroehlichMSS-1985},
simplifications \cite{DreifusK-89} and adaptations to the continuous model on
$L^2(\RR^d)$ \cite{MartinelliH-84,Klopp-95b,CombesH-94b}, which we are considering.
More recently it was used also for Hamiltonians governing the motion of classical waves
\cite{FigotinK-1996,FigotinK-1997a,Stollmann-1998,CombesHT-1999}.

At the same time extensive research has been done to identify physical
situations where one can prove the key ingredients needed to start and
carry trough the MSA \cite{Klopp-95a,BarbarouxCH-1997,KirschSS-1998a,KirschSS-1998b,Veselic-1998,FischerLM-2000}.

\begin{bem}[MSA Hypotheses]
\label{hypotheses}
Let us fix some notation. For points $x \in \RR^d$ in
the configuration space let $ \| x \|_\infty := \sup \{ |x_i| , i = 1,
\dots ,d\}$ denote the sup-norm.  Let  $ \delta > 0 $ be a small constant 
independent of the length scale $ l_k$ and $ \phi_k(x) \in C^2$
a function which is identically equal  to $0$ for $x$ with $ \| x\|_\infty > l_k -\delta$
and identically equal to one for $x$ with $ \| x\|_\infty < l_k - 2 \delta $.
The commutator 
$ W(\phi_k ) := [-\Delta, \phi_k]:= -(\Delta \phi_k) -2  (\nabla \phi_k) \nabla$
is a local operator acting on functions which live on a ring of width $ \delta $
near the boundary of $ \Lambda_k := \Lambda_{l_k} $. We say that a pair 
$ ( \omega,\Lambda_k ) \in \Omega \times {\cal B } ( \RR^d) $ is {\it 
$m$-regular} for a given energy $E$, if
\begin{equation}
\label{ExpDecOnlk}
\sup_{\epsilon \neq 0} \| W (\phi_k ) (H_\omega^l -E + \epsilon i )^{-1}
 \chi^{l_k/3} \|_{{\cal L}(L^2)} \le e^{-ml_k} \ .
\end{equation}
Here $ \chi^{l_k/3} $ is the characteristic function of
$ \Lambda_{l_k/3} := \{ y| \, \| y  \|_\infty \le l_k /6 \} $. Thus the
distance
of the  supports of $ \nabla \phi_k $ and $ \chi^{l_k/3} $ is at least
$ l_k/3 -2\delta \ge l_k/4 $.

There are two key hypotheses for the MSA associated to energies $E$
in the interval $I \subset \RR$ in which one wants to prove the
existence of pure point spectrum.

(H1) $\Leftrightarrow$ There exist constants $ Q_1 \in ]0,\infty [, m \in ]Q_1^{-1} ,\infty [, q > 0$ such that 
\begin{equation}
\PP \{ \omega | \, ( \omega , \Lambda_{Q_1} ) \mbox{ is $m$-regular} \} \ge 1 -
Q_1^{q}.
\end{equation}

(H2) $\Leftrightarrow$ There exist constants $Q_2, \eta_0,  \in
]0,\infty [$ such that 
\begin{equation}
\label{H2}
\PP \{ \omega | \, d( \sigma (H_\omega |_\Lambda ), E ) \le \eta \} 
\le C_W  \: \eta \, | \Lambda |
\end{equation}
for all boxes $\Lambda$ with side length larger than $Q_2$ and all $\eta \le
\eta_0$.
\end{bem}

The first hypothesis (H1) is commonly called {\em initial scale estimate}. 
It provides the induction anchor for the MSA. Most papers  deduce (H1)
from the asymptotic behaviour  of the IDS at so-called spectral 
{\em fluctuation boundaries}. This asymptotics reflect the fact that
``electron levels'' are very sparse near such edges of the spectrum.
The existence of these tails  has been first deduced on physical grounds by Lifshitz \cite{Lifshitz-65}.

The estimate (H2) is associated with a paper of Wegner \cite{Wegner-81}
where he --- like Fr\"ohlich and Spencer \cite{FroehlichS-83} ---
considers Schr\"odinger operators on $l^2(\ZZ^d)$. Wegner's estimate
is needed to draw the induction conclusion on each length scale
$l_k, k \in \NN $ of the MSA.
This is the reason why --- in contrast to (H1) --- it has to be valid
for arbitrarily large scales $l \ge Q_2$ ${}^{1}$.

\footnotetext[1]{Each hypothesis in Remark \ref{hypotheses} has its own
  initial scale: $Q_1$ and $Q_2$. On the other hand the MSA itself
  needs a sufficiently large starting scale $ Q_0$. For the
  whole argument to work out one has to make sure that $Q_1$ can be
  chosen at least as large as the maximum of $Q_0$ and $Q_2$.}
  
Once the MSA has been accomplished, one proceeds to prove localization using the  spectral averaging technique  and expansion in generalized eigenfunctions, cf.~\cite{CombesH-94b} or Sections 7 and 8 in \cite{KirschSS-1998a}. For a different version of the MSA see \cite{Stollmann-2001}.

Actually for the MSA a variety of weaker bounds than (\ref{H2}) is
sufficient. It is enough to know

(H2') $\Leftrightarrow$ There exist constants $Q_2, \eta_0, a,b \in
]0,\infty [$ such that 
\begin{equation}
\label{H2'}
\PP \{ \omega | \, d( \sigma (H_\omega |_\Lambda ), E ) \le \eta \} 
\le C_W  \: \eta^a | \Lambda |^b
\end{equation}
for all boxes $\Lambda$ with side length larger than $Q_2$ and all $\eta \le
\eta_0$.
Inequality (\ref{H2'}) is implied by the H\"older continuity of the averaged finite volume
IDS, mentioned in Remark \ref{questions}: 
\begin{equation}
\label{weak-E}
\EE  \{ N_\bullet^l (E + \eta ) - N_\bullet^l (E - \eta ) \}  
\le C_W  \: \eta^a | \Lambda_l |^{b-1}
\ .
\end{equation}

 We discuss briefly related
results on Wegner estimates and localization for single site potentials with changing sign. 

In \cite{Klopp-95a} Klopp proves a  Wegner estimate like (H2') for the alloy-type 
model (\ref{model})  at low energies. It is valid for energy intervals $[E - \eta, E + \eta] \subset [-\infty , E_c]$ where $E_c$ is an energy strictly above the infimum of the spectrum of $H_\omega$. The result applies  to arbitrary
dimensions $d$, and for the single site potential $u$ only some mild
regularity and decay assumptions are required, but no sign-definiteness.
The density $f$ has to belong to a nice class of functions
which contains as a subset $C^1(\RR)$.
The paper \cite{HislopK-2001} of Hislop and Klopp  improves the volume dependence of the Wegner estimate in \cite{Klopp-95a} using results from \cite{CombesHN-2001} and extends the validity of the estimate to energy values near internal spectral edges.

The techniques developed in \cite{Stolz-2000,BuschmannS-2001}  by Stolz, resp.~Buschmann and Stolz for one-dimensional Schr\"odinger operators  allow to deduce localization at all energies without proving a Wegner estimate. The method  applies to potentials of Poissonian or random
displacement type as well as  to the alloy-type model (\ref{model})  in one dimension with no sign restrictions on the single site potential $u$, cf.~also \cite{DamanikSS-2000}. 

\section{From the finite volume IDS to localized spectral projections}
In this section we reduce the bound of the finite volume IDS to averaging  of
spectral projections localized in
space. We follow the arguments of \cite[Section 4]{CombesH-94b} which
in turn is a generalization of \cite{KotaniS-87}. 
Let $I := ] E_1, E_2[$ be an open energy interval, $P_\omega^l (I)$ the
spectral projection of $H_\omega^l$ onto the interval $I$ and let
$\Tr(A)$ denote the trace of an operator $A$.
Without loss of generality we assume $w = \kappa \chi_{[0,1]^d}$ since only the lower bound matters. Moreover, by rescaling the density $f$ we can achieve $\kappa =1$.
For the finite volume IDS and any $\epsilon > 0$ we have
\begin{equation}
\EE \l [ N_\omega^l (E_2) - N_\omega^l (E_1 + \epsilon) \r ]
\le \frac{1} {l^d} \EE \l [ \Tr P_\omega^l(I) \r ]
\ .
\end{equation}
Let $\tilde{\Lambda} := \Lambda \cap \ZZ^d$ be the lattice points in $\Lambda$. 
As in  \cite{CombesH-94b} we estimate 
\begin{equation}
\label{CH}
\EE \l [ \Tr P_\omega^l(I) \r ]
\le e^{E_2} C_V \sum_{j \in \tilde{\Lambda}} 
\, \l \|  \EE \l [ \chi_j P_\omega^l (I) \chi_j \r ] \r \|
\end{equation}
where  $\chi_j$ is the characteristic function of the unit cube centered at $j$ and the constant $C_V$ is an uniform upper bound on $\Tr (\chi_j e^{-H_\omega^{\Lambda+j}} \chi_j)$, cf.~proof of Theorem 76 in \cite{ReedS-78}.
Here $\Lambda+j$ denotes the unit cube centered at $j \in \ZZ^d$ and $H_\omega^{\Lambda+j}$ the restriction of $H_\omega$ 
on this cube with Neuman b.c. 
For the bound on the operator norm in (\ref{CH}) it
is sufficient to consider $ \EE [ \langle \phi , \chi_j P_\omega^l(I) \chi_j
\phi \rangle ]$ for all normalized  $\phi \in L^2 (\Lambda_l)$.

\section{Transformation of variables}
\label{trans}

In this section we introduce a transformation of variables on the
probability space $\Omega$. It will enable us to use a spectral
averaging result from \cite{CombesH-94b} to bound the expectation
value on the rhs of (\ref{CH}). At the same time we have to keep
control of the new probability density, which will lose its simple 
product structure by the transformation.

Let $A := \{ a_{j,k} \}_{j, k \in \ZZ^d} $ be an infinite Toeplitz matrix
with entries $a_{j,k} = \alpha_{j-k}$. It transforms the components of
the random vector linearly: $\eta:= A \, \omega \in \Omega$.
Note that due to the assumptions on the vector $\alpha$ the matrix $A$
is invertible  and one can derive a bound on its inverse by a Neumann
series.  The row-sum matrix norm we use is given by $
\|A \| := \|A \|_1 := \sup_{j \in \ZZ^d} \sum_{k \in \ZZ^d} | a_{j,k} |
$.
Since $A$ is a Toeplitz matrix generated by the vector $\alpha$ we can
write it as 
\begin{equation}
\label{Neumann}
A =: \Id + S , \  \| S \| = \sum_{j \neq 0} | \alpha_j | = \alpha^* < |\alpha_0|
\end{equation}
by the definition of $\alpha^*$. By rescaling we can assume $\alpha_0 = 1$. So $A^{-1}$ exists and we have 
$\| A^{-1} \| \le \frac{1}{1-\alpha^*}$. Now we introduce truncations 
$A_\Lambda$ of the matrix $A$ associated to a cube $\Lambda \subset \RR^d$. 
Denote with $\Lambda^+ =  \tilde{\Lambda} - \Gamma = \{ \lambda - \gamma | 
\ \lambda \in \tilde{\Lambda}, \gamma \in \Gamma \}$ the set of all sites $k$ 
whose associated potentials $u(\cdot -k)$ influence the potential 
$V_\omega$ within the cube $\Lambda$. The truncated matrix $A_\Lambda = \{
\alpha_{j-k} \}_{j,k \in \Lambda^+}$ acts on $\omega_\Lambda= \{ \omega_k, k \in \Lambda^+ \}$ to
give a vector  $\eta_\Lambda =\{ \eta_k, k \in \Lambda^+ \}$
\begin{equation}
(\eta_\Lambda)_j = (A_\Lambda \omega_\Lambda)_j = \sum_{k \in \Lambda^+} \alpha_{j-k} (\omega_\Lambda)_k 
\ .
\end{equation}
The decomposition and the bound in (\ref{Neumann}) remain true for the
truncations $A_\Lambda$.

We write now the restricted Schr\"odinger operator $H_\omega^l$ in the new
variables $\eta$. We drop for the remainder of this section the index in
the truncated matrix $A_\Lambda$ and vectors  $\omega_\Lambda,\eta_\Lambda$, and write $\tilde{V}_{\eta} $ for $V_\omega =
V_{A^{-1}\eta}$,  $\tilde{H}_{\eta}^l$ for $H_\omega^l = H_{A^{-1} \eta}^l$ and similarly $\tilde{P}_\eta^l = P_\omega^l$ for the operators living on $\Lambda = \Lambda_l$.
For $x \in \Lambda$ we have
\begin{eqnarray}
\nonumber
V_\omega (x)  & = & \sum_{k \in \Lambda^+} \omega_k \sum_{l \in \Gamma} \alpha_l
\, \chi_{k+l} (x)
 =  \sum_{j \in \tilde{\Lambda}} \chi_j(x) \sum_{k \in \Lambda^+} \alpha_{j-k}  \omega_k
\nonumber
\\
& = &
\sum_{j \in \tilde{\Lambda}} \eta_j \chi_j (x)
 = 
\tilde{V}_{\eta} (x)
\ .
\end{eqnarray}
The common density of the random variable $\eta_k, k \in \Lambda^+$ is
given by
\begin{align}
k( \eta) &=  | \det A^{-1}| \, F(A^{-1}\eta), \ 
F(\omega)=\prod_{k \in \Lambda^+} f \l ( \omega_k \r)
\end{align}

\section{Bounds on the density}
\label{Botd}

The structure of the random operator $\tilde{H}_{\eta}^l$ in the new variables makes
it easier to estimate the expectation value 
\begin{equation} 
\label{E}
 \EE \l [ \langle \phi , \chi_j P_\omega^l(I) \chi_j \phi \rangle \r ] 
=\EE \l [ \langle \phi , \chi_j \tilde{P}_\eta^l(I) \chi_j \phi \rangle \r ] .
\end{equation}
We single out the lattice point $j \in \tilde{\Lambda}$ and consider the
one-parameter family of operators
\begin{equation} 
\label{one-parameter}
\eta_j \mapsto \tilde{H}_{\hat \eta}^l (\eta_j)
\ ,
\end{equation}
where $\hat{\eta } = \{ \eta_k , k \in \Lambda^+ \setminus \{ j\} \} $ and 
$\tilde{H}_{\hat \eta}^l (\eta_j) = \tilde{H}_{ \eta}^l$.
Similarly we write $k_{\hat \eta} (\eta_j) = k (\eta)$ for the common density. 
Locally on the cube $\Lambda_1 + j $ the dependence (\ref{one-parameter}) on the parameter $\eta_j$ is
strictly increasing. The price we had to pay is that the random variable $\eta_j$ is (negatively) correlated to  components of $\hat{\eta}$. 
Set $L= \# \Lambda^+$. For a normalized vector $\phi \in L^2 (\Lambda_l)$ set
\begin{equation}
 s(\eta ) := \langle \phi, \chi_j \tilde{P}_\eta^l (I) \chi_j \phi \rangle
\ .
\end{equation}
By Fubini's theorem we have for (\ref{E})
\begin{equation}
\label{sup}
\int_{\RR^L} d\eta \, k( \eta ) \, s( \eta ) 
= \int_{\RR^{L-1}} d\hat{\eta} \,  
\int_{\RR} d\eta_j \, k(\eta ) \, s(\eta ) 
\end{equation}
We bound the rhs of (\ref{sup}) as in \cite[Section 4]{CombesH-94b}.
\begin{equation}
 \int_{\RR} d\eta_j \, k(\eta ) \, s(\eta )
\le 
|I| \ \| k_{\hat \eta}(\cdot)\|_\infty .
\end{equation} 
The bound is valid for $f$ without compact support, too, as pointed out in \cite{FischerHLM-1997}. Furthermore the boundedness condition  on $w$ in  \cite{CombesH-94b} can be replaced with relative boundedness, which is ensured by $w \in L^p$ with appropriate $p$. Note that we have normalized the random potential $V_\omega$ in such a way that the constant $c_0$ from  \cite{CombesH-94b} is equal to $1$.

Using Fubini's theorem in the reverse direction and transforming back
to the $\omega$-variables we estimate (\ref{E})
by
\begin{multline}
 \nonumber
|I| \ \int_{\RR^{L-1}} d\hat{\eta} \,  \| k_{\hat \eta}(\cdot)\|_\infty 
   \le  \nonumber
|I| \ \int_{\RR^{L-1}} d\hat{\eta} \,  \int_\RR d\eta_j \ | k_{\hat \eta}'(\eta_j) |
\\ 
  \le 
|I| \ \int_{\RR^{L}} d\eta \       | k_{\hat \eta}'(\eta_j) |
 =  \label{DerRueckTrnsf}
|I| \, |\det A| \ \int_{\RR^{L}} d\omega \,   | k_{\hat { A \omega }}' \l [(A \omega )_j \r ] |.
\end{multline} 
Let $B = \{ b_{i,j} \}_{i,j\in \Lambda^+}$ denote the inverse of $A = A_\Lambda$.
We  calculate the derivative of $k$ with respect to $\eta_j$
\be
k'_{\hat{ A \omega }} [( A \omega )_j] 
 = 
|\det B|\,  \sum_{k \in \Lambda^+} f' ( \omega_k) \ b_{k, j} \prod_{\iznad{i \in \Lambda^+ }{ i \neq k}} f(\omega_i)
\ee
and  the corresponding integral
\begin{eqnarray}
\int_{\RR} d\omega  \ | k'_{\hat{ A \omega }} [(A \omega )_j] | 
& \le &
|\det B|\,  
\| f' \|_{L^1} \sum_{k \in \Lambda^+} |b_{k,j}|.
\end{eqnarray}
This gives for the expectation (\ref{E}) the estimate
\begin{equation}
\label{main-estimate}
\EE \l [ \langle \phi , \chi_j P_\omega^l(I) \chi_j  \phi \rangle \r ] 
\le 
|I| \ \| f' \|_{L^1} \sum_{k \in \Lambda^+} |b_{k,j}|.
\end{equation}
By the bound (\ref{Neumann}) on the inverse of $A$ we know
$
\sum_{k \in \Lambda^+} |b_{k,j}|    \le \|B \| \le (1-\alpha^*)^{-1}.
$
So (\ref{E}) is bounded by
$ |I| \, \| f' \|_{L^1}  (1-\alpha^*)^{-1} $
which is independent of $ \Lambda_l$ and $j$. The average trace 
in (\ref{CH}) is thus bounded by
\begin{equation}
e^{E_2} C_V  (1-\alpha^*)^{-1} \, \| f' \|_{L^1} \, |I| \, |\Lambda|
\ .
\end{equation}
Thus we proved that the averaged finite volume IDS is Lipschitz continuous
\begin{equation}
\label{fvdos}
\EE \l [ N_\bullet^l (E_2) - N_\bullet^l (E_1 + \epsilon) \r ] 
\le   C \, |E_2- E_1|, \ \forall \ \epsilon > 0
\end{equation}
with $C := e^{E_2} C_V \frac{1}{1-\alpha^*}\, \|f'\|_{L^1}$. By the \v{C}eby\v{s}ev
inequality  estimate (H2) now follows.

\section{Generalizations}

We consider some generalizations of Theorem
\ref{theorem1}. Details can be found in \cite{Veselic-2001}.

\begin{bem}[Discrete model]
One could also consider the discrete Schr\"odinger operator 
\begin{equation}
h_\omega = -\Delta_{\text{disc}} + V_\omega \text{ on } l^2 (\ZZ^d)
\end{equation}
where $[-\Delta_{\text{disc}} \phi](i) = \sum_{|i-n|=1} ( \phi (i) -
\phi (n))$ and in the definition of the multiplication operator $V_\omega$ the
characteristic function of the unit cube $\chi_0(x)$ is replaced by the
Kronecker symbol $\delta_0(i)$. 

Our proof works for this model, too, since the results in
\cite[Section 4]{CombesH-94b} 
are formulated for abstract one-parameter families of operators. In
Sections 3 to 5 of this paper we would just have to change the
notations.

\end{bem}

\begin{bem}[Correlated potentials]

We can regard Theorems \ref{theorem1} as a result about alloy type
Schr\"odinger operators with non-negative single site potential $u$
but negatively correlated coupling constants. I.e.~we consider $\tilde
H_\eta$ as the original operator.

Wegner estimates for dependent coupling constants with bounded conditional densities were derived in \cite{CombesHM-1998} (cf.~also \cite{HupferLMW-2001}).
\nocite{CombesHKN-2001}

Correlated random potentials are also treated in the papers
\cite{DreifusK-91,KirschSS-1998b} on the discrete Anderson model and the alloy type  model. 
There the long range correlations are studied, and the way the MSA has
to be adapted to yield localization in this case.
\end{bem}

\begin{bem}[More general convolution vectors $\alpha$]
The condition $\alpha^* < |\alpha_0|$ in Theorem \ref{theorem1} can be
relaxed as can be seen from the following example.

\begin{bsp}
Let $e = ( 1,0 , \dots, 0) \in \ZZ^d$ and $u = \chi_0 - \chi_e$ be the
single site potential of $H_\omega$ (\ref{model}). This
corresponds to $\alpha_0 = 1, \alpha_e = -1$ and $\alpha_k =0$
otherwise. The truncations $A_\Lambda$ have  inverses $B_\Lambda := \{ b_{j,k} \}_{j,k\in \Lambda^+}$ with entries $ b_{j,k} = 1$ for $j,k \in \Lambda^+$, $k_1 \le j_1$  and $k_i = j_i$ for $i = 2,\dots ,d$ and $b_{j,k} = 0$ otherwise. Here for a $i = 1,\dots , d$ the numbers $k_i$ and $j_i$ denote the $i$-th components of the vectors $k,j \in \ZZ^d$.  In this case the  $B_\Lambda$ are not uniformly bounded in $\Lambda$. However, the term in (\ref{main-estimate}) depending on  $B$ can be estimated by putting all $b_{k,j}=1$:
 \be
  \sum_{k \in \Lambda^+} |b_{k,j}|  \le
 |\Lambda_l^+|
 \ .
 \ee
 Since $| \Lambda_l^+| \le (l +g)^d \le C_\Gamma \,  l^d$, where $g = \diam \Gamma$, we obtain the estimate
 \begin{equation}
 \EE \l [ N_\bullet^l (E_2) - N_\bullet^l (E_1 ) \r ] 
 \le  {\tilde C} \, |E_2- E_1|  \, |\Lambda_l|
 \end{equation}
 where $\tilde C = e^{E_2} C_V C_\Gamma \, \| f'\|_{L^1}  $.
The same Wegner estimate extends to similar single site potentials, e.g.~
\begin{equation}
u = \chi_{(0,\dots,0)} + \chi_{(1,1,0,\dots,0)} 
-\chi_{(1,0,\dots,0)} - \chi_{(0,1,0,\dots,0)} \, .
\end{equation} 
\end{bsp}
Note  that while we have proven (\ref{weak-E}) with the exponent
$b=2$  we cannot deduce the Lipschitz continuity of the IDS
because of the divergent term $|\Lambda_l|^2$. This example illustrates that
(H2') can be proven for $A$ (respectively $\alpha$) with
inverses whose norms grow at most polynomially in $|\Lambda|$. It would be
desirable to get a nice description of this class of Toeplitz
matrices in terms of the convolution vector $\alpha$, cf.~Remark \ref{Serra}.
\end{bem}

There is a completely different $f$ then the differentiable densities considered so far we can cope with, namely the
uniform density, as the following example shows.

\begin{bsp}[Uniform density]
Consider again the single site potential $u = \chi_0 -\chi_e$ but now
with the uniform density $f(x) = \frac{1}{\omega_+} \chi_{[0,\omega_+]}$ for
the coupling constants $\omega_k, k \in \ZZ^d$. The reasoning of Sections
3 and 4 remain valid for this case, too. Section 5 has to be replaced
by explicit estimates on the volume of the integration domain $M$ and
the common density $k$ of the transformed variables using 
$ |\supp f| \, \|f\|_\infty = 1$ to get
\begin{equation}
 \EE \l [ N_\bullet^l (E_2) - N_\bullet^l (E_1) \r ] 
 \le  \text{ const } |E_2- E_1| \, |\Lambda_l|
\ .
\end{equation}
See \cite{Veselic-2001} for the details and \cite{KostrykinV-2001} for extensions.
Unfortunately we cannot deal with the superposition of the uniform and
$W_1^1$-densities due to the transformation $A^{-1}$ which appears in
the common density $k$.
\end{bsp}

\section{Localization}
\label{sec-localization}
An important application of Wegner's estimate is the proof of
localization. This raises the question whether there is a class
of single site potentials $u$ for which Theorem \ref{theorem1} is valid
and additionally an initial scale estimate (H1) can be proven. As
mentioned before, for non-negative $u$, Lifshitz tails can be used to
deduce (H1) near the infimum of the spectrum of $H_\omega$. Now, for $u$
with changing sign there are only restrictive results on Lifshitz asymptotics (cf.~Section 6.2.2 in \cite{HislopK-2001}).
Most proofs are not stable under a (even small) negative perturbation of $u$.
However, while the standard deduction of the asymptotic behaviour is
based on a sequence of inequalities for the first Neumann eigenvalue of
$H_\omega^l$ on arbitrarily large cubes $\Lambda_l$,  (H1) is implied by this inequality
on a sufficiently large, but {\em fixed} $\Lambda_{Q_1}$.

We will show that the basic estimate on the first Neumann eigenvalue
on a fixed scale $\Lambda_{Q_1}$ is stable under a negative perturbation of
$u$ as long as it is coupled with a small parameter $\epsilon_u$. The dependence
(see also Remark \ref{epsilon-Q}) $\epsilon_u = \epsilon_u (Q_1) \to 0$
for  $Q_1 \to \infty$ explains why our estimate is no good for proving
Lifshitz tails.

Throughout this section we assume that the support of $f$ is an bounded interval. By changing the periodic potential
we can assume $\supp f = [0, \omega_+]$. Again, details of the proofs can be found in \cite{Veselic-2001}.
\begin{nta}[Small negative perturbation of $u$]
We decompose $u = u_+ - \epsilon_u u_-$ into a non-negative $u_+$ and a
non-positive part $-\epsilon_u u_-$, with $\|u_-\|_\infty \le 1,
\epsilon_u \in [0,1]$ and $\supp u \subset \Lambda_g, g> 0$. We set 
$N = \| \sum_{k \in \ZZ^d} u_- ( \cdot - k) \|_\infty $.

The following arguments are adaptations of
inequality (2) and Proposition 3 in \cite{KirschS-86} to $u$ with
changing sign.
The restriction of $H_\omega$ and $H_0= -\Delta + V_0$ to $\Lambda_l$ with Neumann b.c.~will
be denoted by $H_\omega^{l,N}$ and $H_0^{l,N}$ respectively. 
Assume that $V_0$ is symmetric under the
reflection along the coordinate axes. Let $\phi$ be the ground state
of $H_0^{1,N}$  and $\Phi$ its periodic extension to $\RR^d$. Then for
$ l \in \NN$, $ |\Lambda_l|^{-1/2} \, \Phi \, \chi_{\Lambda_l}$  is the ground
state of both $H_\omega^{l,N}$ and $H_\omega^{l,\text{per}}$, where ``per'' stands for
periodic b.c. So we have 
\begin{equation}
\inf \sigma  (  H_0 )  = \lambda_1 \l ( H_0^{1,\text{per}} \r )  
= \lambda_1 \l ( H_0^{l,\text{per}} \r )  = \lambda_1 \l ( H_0^{l,N}
\r )  
\ .
\end{equation}
By adding a constant we get
\begin{equation}
\inf \sigma  (  H_0 )  =0
\ .
\end{equation}
Set $m_1 = \int dx \,u(x) \Phi^2 (x) $ and assume that $\epsilon_u$ is
so small that $m_1 > 0$. For a given energy $E \in ]0,1[$ and a parameter
$\beta > 0$ choose the length scale $ l := [ (\beta E )^{-1/2} ] $.
\end{nta}
By Dirichlet-Neumann bracketing we know
\begin{equation}
\label{D-N-b}
\PP \{ \sigma ( H_\omega^{l} ) \, \cap  \, ] -\infty ,E[   \, \neq \emptyset  \}
\le
\PP \{ \omega |  \, \lambda_1( H_\omega^{l,N} ) < E \} \, ,
\end{equation}
where $H_\omega^{l} $ may have periodic, Dirichlet or Neumann b.c.
 We will derive
an upper bound on $\PP \{ \omega | \lambda_1( H_\omega^{l,N} ) < E \} $
which is exponentially small in $|\Lambda_l|=l^d$. The exponential bound
follows from the combination of a Large Deviations result and the fact
that $\lambda_1 (H_\omega^{l,N})$ can attain a small value only for very
rare configurations of $\omega$.

\begin{prp}
\label{rare-config}
There exist $ \beta_0, Q_1 < \infty $,  such that for
$ l \ge Q_1, \beta \ge \beta_0 $ and $ \epsilon_u \le \frac{E}{8 \omega_+ N} $
we have the estimate:
\bea
\lambda_1  \l (  H_\omega^l  \r )  < E \Longrightarrow  \;
\#  \l \{ k \in \Lambda_l  \l |  \,  \omega_k <  \frac {4E}{m_1} \r . \r \}
> \frac {l^d}{2} \, .
\eea
\end{prp}
The proof is an adaptation of the one  of \cite[Proposition 3]{KirschS-87} and can be found in \cite{Veselic-2001}. One has just to control the contributions from $u_-$ and is not allowed to
replace $u$ by $u \chi_0$ as done in \cite{KirschS-87}.

By Large Deviations we know
\begin{equation}
\label{LD}
\PP \l \{ \#  \l \{ k \in \Lambda_l  \l |  \,  \omega_k < 4E/m_1 \r . \r \}
> \frac {l^d}{2} \r \} \le  e^{-c|\Lambda_l|}
\end{equation}
where we choose $E$ sufficiently small so that $\EE (\omega_0) > \frac{4
  E}{m_1}$. $c >0$ is a  constant independent of $l$. Combining
(\ref{D-N-b}), Proposition \ref{rare-config} and (\ref{LD}) we arrive
at the bound
\begin{equation}
\label{small-prob}
\PP \{ \sigma ( H_\omega^{l} )   \, \cap  \, ] -\infty ,E[  \,  \neq   \, \emptyset  \}
\le    e^{-cl^d} \, .
\end{equation}
The relation $ l \approx E^{-1/2} $ is not appropriate  for the deduction of property (H1), so we have to introduce an second length scale $L$. 
Consider he operator $H_\omega^L$ on a larger cube $\Lambda_L$ which is split by Neumann surfaces into cubes with side length 
$l := [ L^{1-\zeta/2} \beta^{-1/2} -1], \, \zeta \in ]0,1[$.
The operator on the cube $\Lambda_l +j$ for $j \in (l \ZZ)^d \cap \Lambda_L$ is denoted by $H_{\omega,j}$. 
We have
    \[
    \lambda_1 (H_\omega^L) 
    \ge \inf_{\textstyle j \in (l \ZZ)^d \cap \Lambda_L} \lambda_1 (H_{\omega,j}) .    
    \]
and thus using (\ref{small-prob})  
\begin{multline}
    \label{l-exp-klein}
    \PP\{ \, \lambda_1 ( H_\omega^{L,N}) < L^{-2 +\zeta } \}
        \le
    \PP\{ \, \lambda_1 ( H_\omega^{L,N}) \le \beta^{-1} ( l +1 )^{-2} \}
    \\
    \le
    \PP \l \{ \, \inf_{j \in (l \ZZ)^d \cap \Lambda_L} \lambda_1 ( H_{\omega,j}) \le \beta^{-1} ( l +1 )^{-2}  \r \}
    \\
        \le
    \sum_{ j \in (l \ZZ)^d \cap \Lambda_L}
    \PP\{ \,  \lambda_1 ( H_{\omega,j}) \le \beta^{-1} ( l +1 )^{-2} \}
    \\
    \le
    \l ( \frac{L}{l} \r )^d
    \PP\{ \,  \lambda_1 ( H_{\omega,0}) \le \beta^{-1} ( l +1 )^{-2} \}
    \\
     \le
    \l ( \frac{L}{l} \r )^d e^{-\l (\frac{11}{12} \r )^2\frac{l^d}{2}}
    \le L^{-q}
    \end{multline}
    for any $\zeta \in ]0,1[$ and $q \in \NN $ for $L$ large enough.

  Note that due to the condition in Proposition \ref{rare-config} the last inequality is applicable for single site potentials $u= u_+-\epsilon_u u_-$ with $ \epsilon_u \le (8 \omega_+ N  \beta (l+1)^2)^{-1} $.
Applying the Combes-Thomas argument (cf.~\cite{CombesT-73}, \cite{BarbarouxCH-1997}, \cite[Appendix]{KirschSS-1998a} 
or \cite[Section 2.4]{Stollmann-2001}) the
initial scale estimate (H1) follows.

Note that by
\cite[equation (1.1)]{KirschSS-1998a} $\sigma ( H_\omega) \supset \sigma
(H_0) \ni 0$. This implies that for any $E>0$ the interval $[0,E]$ actually does contain spectrum. Now localization at the bottom of the spectrum follows. 

\begin{thm}
\label{localization}
Let $H_\omega$ be as in Theorem \ref{theorem1}. Assume that $V_0$ is symmetric under the
reflection along the coordinate axes and 
$\supp f = [0,\omega_+]$. Then there exist $\epsilon_u > 0$ and $E^* > 0 = \inf
\sigma (H_0)$ such that sufficiently small $\epsilon_u$: 
\begin{equation}
\sigma (H_\omega) \, \cap \, ]-\infty , E^*[  \neq \emptyset , \  
\sigma_{c} (H_\omega) \, \cap \, ]-\infty , E^*[ = \emptyset  .
\end{equation}
\end{thm}

We turn our attention now to localization away from the infimum of
$\sigma(H_\omega)$. The spectrum of the periodic Schr\"odinger operator
$H_0$ consists of intervals called spectral bands. If there are gaps
belonging to the resolvent set  between them, there exist internal
spectral (band) edges. If the perturbation  $V_\omega$ is small, the
spectral gaps will be preserved, although  with shifted
spectral edges. It is natural to ask whether the localization results
for energies near $\inf \sigma (H_\omega)$ can be extended to small
neighbourhoods of internal edges. It turns out that the study of
Lifshitz tails in this energy regime is quite involved
\cite{Klopp-1999}.

As a substitute for the Lifshitz asymptotic a special disorder regime
has been assumed in several papers
\cite{BarbarouxCH-1997,KirschSS-1998a}. In this case the density $f$
of the coupling constants is required to satisfy
\begin{equation}
\label{polynomial-f}
\int_{\omega_-}^{\omega_- + \delta} f(x) \,  dx \le \delta^\tau \text{ or } 
\int^{\omega_+}_{\omega_+ - \delta} f(x) \, 
\le \delta^\tau \text{ for some $\tau > d/2$ and small $\delta$.}
\end{equation}
The first condition is needed when considering lower band edges, the
second for upper band edges.
In this case we can prove using the arguments of
\cite[pp.10-11]{KirschSS-1998a}:

\begin{prp}
\label{prp-ini-indef}
Let $H_\omega,f$ be as in Theorem \ref{localization},
let $f$ satisfy (\ref{polynomial-f}) and $E$ be  a  spectral band
edge. 
Let $p \in ]0,2 \tau -d[$ and $\xi \in ]0,2-\frac{d+p}{\tau}[$. Then there exists a
$Q_1$ such that for all $l \ge Q_1$ and $ \epsilon_u \le \frac{l^{\xi-2}}{\omega_+N}$ we have
\begin{equation}
 \PP \{ \sigma (H_\omega^{l,\mbox{per}}) \cap  [
E-l^{\xi-2}, E+ l^{\xi-2}] \neq \emptyset \} \le l^{-p} \, .
\end{equation}
\end{prp}

Now one proceeds as in \cite{BarbarouxCH-1997,KirschSS-1998a} or \cite{Stollmann-2001} using Combes-Thomas
arguments to prove an initial scale estimate and thereby
localization. 
\begin{thm}
\label{int-loc}
Let $H_\omega$ and $f$ be as in Proposition \ref{prp-ini-indef} and  let be $E$  a spectral band
edge.
There exist  $r> 0$ such that for sufficiently small $\epsilon_u $ the spectrum of $H_\omega$ in the
interval $[E-r, E+r]$ is pure point.
\end{thm}
In Section 6.2 of \cite{HislopK-2001} it is described how to prove this result  for a larger class of
 single site potentials and density functions $f$,  using results from \cite{Klopp-1999} and an abstract version of the smallness of $u_-$. To apply this reasoning it is necessary that the unperturbed periodic operator $-\Delta+V_0$ is non-degenerate (resp.~Floquet-regular) at the considered spectral boundary.  

\begin{bem}
\label{epsilon-Q}
For the localization Theorems \ref{localization}  and \ref{int-loc} we had to choose
$\epsilon_u$ small depending on the initial scale $Q_1$. It might seem
irritating that $Q_1$ in turn depends on $V_\omega$, i.e.~implicitly on $\epsilon_u$.
However an admissible initial scale $Q_1$ for $\epsilon_u =1$, i.e.~the potential
$V_\omega^{\epsilon_u=1}(x) = \sum_{k \in \ZZ^d} \omega_k \, (u_+ -u_-) (x-k)$
is admissible for the potentials for all values of $\epsilon_u \in
[0,1]$, also. This means that choosing $\epsilon_u$
closer to $0$ does not change the initial scale $Q_1$.
\end{bem}


\def\cprime{$'$} \def\cprime{$'$}

\end{document}